\documentclass[
reprint,
superscriptaddress,
amsmath,amssymb,
aps,
pra,
]{revtex4-2}

\usepackage{graphicx}
\usepackage{epstopdf}

\usepackage{dcolumn}
\usepackage{bm}
\usepackage{hyperref}
\usepackage[dvipsnames]{xcolor}
\usepackage{framed}
\usepackage{tcolorbox}
\usepackage{tikz}
\usepackage{xcolor}
\usepackage{multirow}
\usepackage{makecell}
\usepackage{placeins}
\usepackage{mhchem}
\usetikzlibrary{positioning,calc,arrows.meta,shapes.geometric,shapes.multipart,fit}

\def  \loss  {\mathcal{L}}
\def  \bS    {{\bm S}}

\def \lossctr {{\loss}}
\def \losssoft{{\loss_{\text{soft}}}}

\def \op {{\bm \chi}}
\def \sl {{s_{ij}}}

\def \bp {{\bm P}}
\def \bA {{\bm A}}
\def \bX {{\bm X}}
\def \bQ {{\bm Q}}
\def \br {{\bm r}}
\def \bq {{\bm Q}}

\newcommand{\str}[1]{{\overline{\bS_{#1}}}}
\newcommand{\its}[1]{{\overline{I_{#1}}}}

\def \ccor {{\overline{C}}}

\begin{document}

\title{Soft Contrastive Learning for Unsupervised Discovery of Phases of Matter}

\author{Vít Beneš}
\affiliation{FZU - Institute of Physics of the Czech Academy of Sciences, Na Slovance 1999/2, 182 00 Prague, Czech Republic}
\affiliation{Department of Materials, Faculty of Nuclear Sciences and Physical Engineering, Czech Technical University in Prague, Trojanova 13, Prague 120 00,
Czech Republic}

\author{Pavel Baláž}
\affiliation{FZU - Institute of Physics of the Czech Academy of Sciences, Na Slovance 1999/2, 182 00 Prague, Czech Republic}

\author{Dmytro Bohdanov}
\affiliation{FZU - Institute of Physics of the Czech Academy of Sciences, Na Slovance 1999/2, 182 00 Prague, Czech Republic}
\affiliation{Department of Solid State Physics, Faculty of Nuclear Sciences and Physical Engineering, Czech Technical University in Prague, Trojanova 13, Prague 120 00,
Czech Republic}

\author{Jiří Hlinka}
\affiliation{FZU - Institute of Physics of the Czech Academy of Sciences, Na Slovance 1999/2, 182 00 Prague, Czech Republic}

\begin{abstract}
A machine learning (ML) framework for phase-diagram construction in condensed matter systems with unknown phase structure is presented. The method combines physically motivated descriptors with contrastive representation learning and a two-stage data-generation workflow. 
First, a sparse set of configurations is generated by Monte Carlo simulated annealing in order to train a contrastive neural network that produces a low-dimensional embedding. Consequently, representative phase configurations are used as initial conditions for gradient-based optimization on a dense parameter grid, enabling efficient generation of refined datasets for high-resolution phase-diagram construction.
In order to optimize the workflow, a variant of the method, in which derived descriptors are used as soft labels in the contrastive objective, is examined as well. 
The framework is demonstrated on a \ce{PbZrO3}-inspired model potential formulated in terms of coupled order-parameter fields and used as a physically motivated testbed for exploring configuration diversity and phase behavior. 
The proposed approach provides a physically informed route to ML-assisted phase-diagram construction that integrates data analysis with adaptive dataset refinement.
\end{abstract}

\maketitle

\section{Introduction}
In condensed matter and statistical physics, many-body systems produce complex configuration spaces where relevant 
collective behavior is not always captured by simple predefined observables. Traditional analysis of phase transitions relies on identifying suitable order parameters or symmetry-breaking mechanisms. However, in complex or topological systems, 
such quantities may not be known {\em a priori}, or may be difficult to define. Machine learning (ML) has been explored as a practical way forward. Its flexibility allows it to extract salient information from high-dimensional datasets that are difficult to analyze using conventional approaches which require explicit model of the transition. It has consequently become a powerful tool for analyzing phases of matter \cite{carleo_machine_2019,Dawid_2025, Wang_PRB_2016, iakovlev_supervised_2018,frk_unsupervised_2025,mezera_neural_2023,vanNieuwenburg_NatPhys_2017,richter_PRE_2023,arnold_PRRes_2021,arnold_PRX_2022}.

ML methods for phase classification have largely followed two strategies:
supervised classification using predefined phase
labels~\cite{iakovlev_supervised_2018,mezera_neural_2023}, or unsupervised learning
that requires no labels at
all~\cite{Wang_PRB_2016,frk_unsupervised_2025,Hu_PRE_2017,Wetzel_PRE_2017}.
Supervised strategies presuppose knowledge of the
phases that may appear in the data. This assumption is often violated when studying a
new physical system, where we usually do not know how many phases exist or what they
look like. Unsupervised methods avoid this requirement, learning features that
characterize distinct regimes and producing low-dimensional representations that can
be clustered into physical phases. 
Typically, an effective low-dimensional representation can be constructed using 
principal components analysis (PCA)~\cite{Wang_PRB_2016} or autoencoders~\cite{Hu_PRE_2017}.
However, if unsupervised methods are applied directly to raw microscopic configurations the
learned representation may be dominated by microscopic fluctuations rather than by
the collective features relevant for phase
identification~\cite{Hu_PRE_2017,Wetzel_PRE_2017,carleo_machine_2019}.

In many physical systems, one can compute a set of generic descriptors,
such as structure factors or correlations, that capture essential collective
features without being tailored to a specific phase transition. Although these
quantities do not necessarily constitute universal order parameters, they provide a
physically informed and compact representation of the system, and they induce a
notion of similarity between configurations that requires no prior knowledge of the
phases. Their direct use as a feature space is nonetheless restrictive: the
descriptor space is constructed by hand, and distances within it need not reflect
the grouping of configurations into phases. This motivates treating the descriptors
as a training signal rather than as the final representation, and learning an
embedding in which the induced similarity is enforced. Contrastive objectives are
designed for precisely this setting, as they are formulated directly in terms of
pairwise similarity between samples.

Contrastive learning has been widely successful in vision~\cite{chen_simple_2020,he_2020_momentum_contrast,radford_2021_learning_transferable}, speech~\cite{oord_2019_representation_learning, baevski_2020_wav2vec20}, and natural
language processing~\cite{gao_2022_simcse}, and has recently been adopted for phase recognition in
statistical physics by Han \emph{et al.}~\cite{han_framework_2023}, who employed a contrastive strategy on Monte Carlo configurations by defining positives as
configurations obtained at identical control parameters, thereby injecting physical information through the pairing rule rather than through explicit data
augmentations. Here we instead inject physical information through the descriptors themselves, which allows similarity to be defined between configurations at {\em different} control parameters.

Building on this idea, we make three contributions. First, we combine physically
informed descriptors with raw microscopic configurations to improve the quality and
robustness of representation learning. Second, we train a neural network that
produces a meaningful low-dimensional embedding of system states, capturing
phase-relevant similarity beyond what is explicitly encoded in the descriptor space.
Third, we integrate this learned representation into a two-stage data-generation
workflow that exploits the complementary strengths of two sampling strategies; Monte Carlo (MC) simulated annealing and a deterministic relaxation.  
MC simulated annealing requires no prior knowledge of the target state and can start
from a random configuration, but it is computationally expensive and, because the
energy landscape hosts multiple nearly degenerate minima, may terminate in a
higher-energy state. Deterministic gradient relaxation is by contrast fast but
inherently local and requires a sufficiently good initial ansatz to avoid
converging to an undesired metastable minimum. By using representative
configurations identified in the first stage as initial conditions for the second,
our framework couples analysis with adaptive dataset refinement, enabling the
construction of phase diagrams at high resolution while reducing the overall
computational cost.

We demonstrate the proposed framework on a model potential inspired by the
antiferroelectric perovskite \ce{PbZrO3}
(PZO)~\cite{liu_a_brief_review_2011,reyes-lillo_antiferroelectricity_2013,tagantsev_origin_2013,hlinka_2014:PRL}.
Rather than aiming at a faithful description of the material, we use this potential
as a physically motivated testbed to explore the range of possible configurations
and the associated phase behavior across the relevant parameter space. This setting
provides a controlled environment for assessing the ability of our
descriptor-enhanced contrastive approach to discover phases and construct phase
diagrams, while leaving material-specific refinements of the model potential for
future work.

The paper is organized as follows. In Sec.~\ref{Sec:PZO} we introduce the
PZO-motivated potential and describe the data generation and the descriptors used.
Section~\ref{Sec:ML} describes the workflow and the ML model, including the
contrastive-learning method. Section~\ref{Sec:Results} presents the results of our
analysis and the phase-diagram construction, which are discussed in detail in
Sec.~\ref{Sec:Discussion}. We conclude with a summary in Sec.~\ref{Sec:Summary}.

\section{PZO-motivated Potential}
\label{Sec:PZO}

To test our model, we have chosen a physically relevant example of an 
effective interatomic potential motivated by the structure of \ce{PbZrO3} (PZO) --
a perovskite oxide of the ABO$_3$ type.
It is widely regarded as the prototypical antiferroelectric (AFE) material 
\cite{shirane_dielectric_1951, randall_antiferroelectrics_2021}.
On cooling, bulk PZO undergoes a first-order transition at $T_{\rm C} \simeq 500\, {\rm K}$
from the cubic paraelectric perovskite phase (space group {\em Pm$\bar{3}$m}) 
into an orthorhombic antiferroelectric phase ({\em Pbam}), 
characterized by antipolar Pb displacements accompanied by oxygen-octahedra rotations \cite{tagantsev_origin_2013}.
Classic x-ray and neutron diffraction work established the 
low-temperature antiferroelectric structure \cite{jona_x-ray_1957}.
Lattice-dynamical studies using x-ray scattering and Brillouin light scattering 
suggest that the AFE state originates from a single ferroelectric instability 
with an important role of flexoelectric coupling \cite{tagantsev_origin_2013}.
On the other hand, IR and Raman spectroscopy study demonstrate the existence 
a multiple soft modes \cite{hlinka_2014:PRL}. It also suggests that 
AFE states is stabilized by a trilinear term, that connects
the AFE lead displacements to oxygen octahedra tilts. 
An effective 1-dimensional model of PZO potential including a soft branch part
related to Pb ions displacement and term associated with
oxygen octahedra tilting including the trilinear coupling
leads to a phase diagram demonstrating existence of paraelectric (PE),
ferroelectric (FE), and antiferroelectric (AFE) phase \cite{hlinka_2014:PRL}. 

\subsection{2D effective potential}

Here, we extend the previously suggested 1D potential \cite{hlinka_2014:PRL} 
to study phase transitions in more realistic case of 2-dimensional PZO lattice.
In analogy to the 1-dimensional case, the potential contains {\em soft} (lead ions) and
{\em hard} (oxygen ions) potential branches and their mutual coupling.
The model described two coupled vector fields on a regular 2D square lattice
\begin{subequations}
    \begin{align}
        P_{m, n} &= (u_{m,n}, v_{m,n})_{\rm pc}\,, \label{Eq:pmn} \\
        A_{m, n} &= (\eta_{m,n}, \xi_{m,n})_{\rm pc}\,, \label{Eq:amn}
    \end{align}
\label{Eq:variables}
\end{subequations}
where pc stands for conventional pseudocubic
cartesian system and pair of integers $(m, n)$ label the lattice sites.
$P_{m, n}$ is the in-plane polarization vector (lead ions) while
$A_{m, n}$ stands for a {\em hidden} field of in-plane vectors related
to oxygen octahedra tilting. We denote the pair of fields over the whole lattice as $\bX = (\bp, \bA)$.
For convenience, the potential is described  using
auxiliary field coordinates 
$r_{m,n}$, $s_{m,n}$, $\phi_{m,n}$, $\psi_{m,n}$
referring to an
rotated cartesian system with axes along $\left[\bar{1}1\right]$ and
$\left[11\right]$ as follows
\begin{align*}
    v_{m, n} &= \frac{s_{m,n} + r_{m,n}}{\sqrt{2}}\,, & 
    u_{m,n} &= \frac{s_{m,n} - r_{m,n}}{\sqrt{2}}\,, \\
    \xi_{m,n} &= \frac{\psi_{m,n} + \phi_{m,n}}{\sqrt{2}}\,, & 
    \eta_{m,n} &= \frac{\psi_{m,n} - \phi_{m,n}}{\sqrt{2}}\,.
\end{align*}

The potential under consideration is defined as
\begin{equation}
    V = \sum_{(m, n)} \biggl[
    V_{\rm soft} + V'_{\rm soft} + V_{\rm hard} + V'_{\rm hard} + V_{\rm int}
    \biggr]\,.
\label{Eq:potential}
\end{equation}
The first term stands for the {\em soft} local potential for polarization field
\begin{equation}
    \begin{split}
        V_{\rm soft} =\;
        &\frac{\alpha}{2} \left( r^2_{m,n} + s^2_{m,n} \right) + 
        \frac{1}{4} \left( r^2_{m,n} + s^2_{m,n} \right)^2 + \\
        &\frac{1}{2}\, r^2_{m,n}\, s^2_{m,n}\,,
    \end{split}
\label{Eq:V_soft}
\end{equation}
where $\alpha = a\, (T - T_0)$ with $T_0$ being a critical temperature
for the PE -- FE phase transition.
$V_{\rm soft}$ favors the oblique polarization directions, 
while the {\em hard} local potential for hidden field has the opposite anisotropy
\begin{equation}
    \begin{split}
        V_{\rm hard} =\;
        &\frac{1}{2} \left( \phi^2_{m,n} + \psi^2_{m,n} \right) + 
        \frac{1}{4} \left( \phi^2_{m,n} + \psi^2_{m,n} \right)^2 - \\
        &\frac{1}{2}\, \phi^2_{m,n}\, \psi^2_{m,n}\,.
    \end{split}
\label{Eq:V_hard}
\end{equation}

The dispersion energy for polarization field is defined
by nearest-neighbor (NN) and next-nearest-neighbour (NNN) harmonic interactions
{
\begin{equation}
    \begin{split}
        V'_{\rm soft} =\;
        &\frac{g_1}{2} \biggl[ 
        \left(v_{m,n} - v_{m+1,n} \right)^2 + 
        \left(u_{m,n} - u_{m,n+1} \right)^2  \biggr] + \\
        &\frac{g_2}{2} \biggl[ 
        \left(u_{m,n} - u_{m+1,n} \right)^2 + 
        \left(v_{m,n} - v_{m,n+1} \right)^2  \biggr] + \\
        &\frac{g_4}{2} \biggl[ 
        \left(s_{m,n} - s_{m+1,n+1} \right)^2 + 
        \left(r_{m+1,n} - r_{m,n+1} \right)^2  \biggr]\,,
    \end{split}
\label{Eq:V_soft_prime}
\end{equation}
}
where the coefficients were fixed to favor 
off-axis transverse polarization waves
$g_1 = 1$, $g_2 = 5$, and $g_4 = 10$.

In case of the {\em hidden} field, only the diagonal 
longitudinal fluctuations are suppressed by the NNN harmonic interaction
\begin{equation}
    \begin{split}
        V'_{\rm hard} =                 
        \frac{\lambda_4}{2} \biggl[ 
        &\left(\psi_{m,n} - \psi_{m+1,n+1} \right)^2 + \\
        &\left(\phi_{m+1,n} - \phi_{m,n+1} \right)^2  \biggr]
    \end{split}
\label{Eq:V_hard_prime}
\end{equation}
with $\lambda_4 = 10$.

Finally, the trilinear coupling term reads
\begin{equation}
    \begin{split}
    V_{\rm int} =
    &\frac{\kappa}{2}\, 
    \left(r_{m,n} - r_{m+1, n+1} + 
    s_{m+1, n} - s_{m, n+1} \right) \times\\
    &\phi_{m,n}\, \psi_{m, n}
    \end{split}
\label{Eq:V_int}
\end{equation}
where $\kappa > 0$ is the interaction constant.

The potential $V$ defined by Eqs.~(\ref{Eq:potential}--\ref{Eq:V_int}) is the model studied in this work.
Although the potential depends on several parameters tailoring its shape, we fix most of them and focus on two control parameters, $\alpha$ and $\kappa$.
Thus, the $(\alpha,\kappa)$ plane defines the two-dimensional parameter space 
in which we construct the phase diagram.

\section{Methods}
\label{Sec:ML}

\begin{figure*}[t]
    \centering
    \includegraphics[width=\textwidth]{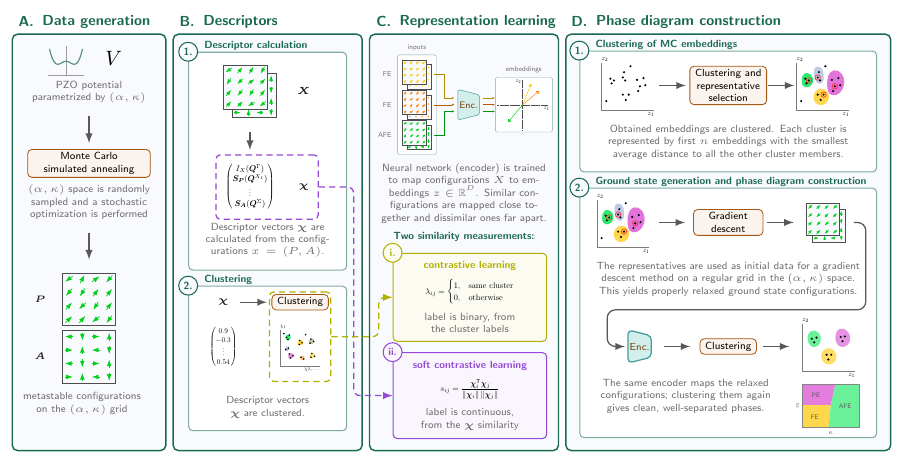}
    \caption{Workflow overview.
    {(A) Training data generation:} MC simulated annealing generates
    metastable configurations across the $(\alpha,\kappa)$ of the PZO potential.
    {(B) Descriptors:} a descriptor vector $\op$ is
    computed per configuration. In case of classical contrastive learning $\op$ is used for a preliminary clustering and construction of positive and negative pairs. In case of soft contrastive learning $\op$ is used to define similarity between two samples.
    {(C) Representation learning:} an encoder is trained via
    contrastive learning, using either cluster membership or distance between descriptor vectors as similarity label.
    {(D) Phase diagram construction:} clustering the MC embeddings and
    selecting representatives yields a small set of
    configurations, which seed a gradient descent on a dense $(\alpha,\kappa)$
    grid.  Mapping the resulting relaxed configurations through the same encoder
    and clustering once more gives the final phase diagram.}
    \label{fig:diagram}
\end{figure*}
The whole workflow can be described in four steps which will be followed throughout the Methods section.
These steps are depicted in Fig.\,\ref{fig:diagram}: 
{(A) Data generation} (Configurations $\bX$ for different $\alpha$ and $\kappa$ parameters of the PZO potential are generated from random initial conditions using the MC simulated annealing algorithm as described in Sec.~\ref{sec:data_gen}.)
{(B) Descriptors calculation and clustering} (Descriptor vectors $\op$ are calculated from the generated configurations. The descriptor vectors are chosen to carry physical information of the system. Clusters of the descriptor vectors are found using a density-based clustering algorithm as described in Sec.~\ref{sec:descriptor}.)
{(C) Representation learning} (Neural network (encoder) which maps configurations $\bm X$ to vectors (embeddings) $z \in \mathbb{R}^D$ for some $D \in \mathbb{N}$ is trained using a contrastive learning paradigm, using information from both the configurations and the descriptors as described in Sec.~\ref{sec:encoder}.)
{(D) Phase diagram construction} ((1) The training configurations are mapped into the embedding space and subsequently clustered, refining the cluster memberships from step B. For each cluster a set of $n$ representatives is identified. (2) When constructing phase diagram, we find ground state for each pair of $(\alpha, \kappa)$ using gradient descent method with the configurations corresponding to found representatives as initial conditions. The optimized configuration with lowest energy is mapped using the encoder. Finally, the ground state configurations are clustered to obtain the phase diagram as described in Sec.~\ref{sec:phaseid}.)
 
In the following, we will describe each stage in more detail.

\subsection{Training data generation} \label{sec:data_gen}

In stage A we generate the training dataset that shall be used to train
the unsupervised ML methods. 
The training dataset consists of low energy configurations of $\bX = ({\bm P}, {\bm A})$
for randomly distributed pairs of $(\alpha, \kappa)$ parameters in the given parameter space.
In order to get rid of any assumption of the configuration, we minimize the potential $V$ [Eq.~(\ref{Eq:potential})] use MC simulated annealing.
In contrast to local gradient descent minimization, this stochastic procedure is less sensitive to the initial condition and helps to escape from shallow metastable states.

We start from a random initial configuration in which
\begin{equation*}
\begin{split}
u_{m,n},\, v_{m,n},\, \eta_{m,n},\, \xi_{m,n} &\sim \mathcal{U}([-1,1])\,,\\
&\forall\, m,n \in \{1,2,\dots,N\}\,,
\end{split}
\end{equation*}
where $\mathcal{U}([-1,1])$ denotes the uniform probability distribution on the interval $[-1,1]$
and $N$ is the linear size of the 2D lattice.
We perform standard Metropolis--Hastings updates while lowering the temperature $\tau$ according to a prescribed cooling schedule.
We optimize both, polarization as well as hidden field variables.
Thus, a configuration of the 2D system
consists of $4 \times N \times N$ independent variables.

The goal of this procedure is not to sample the finite-temperature equilibrium 
distribution, but to obtain candidate ground-state configurations by driving the 
system towards low-energy minima. Note that the physical temperature $T$ enters 
the potential only through the coefficient $\alpha$, at the mean-field level. 
The fictitious temperature $\tau$ used in the simulated annealing controls the 
acceptance of trial moves and is an optimization parameter unrelated to $T$; since 
it is lowered to zero at the end of the schedule, the resulting configurations 
carry no thermal fluctuations. 

In our settings, we chose $N = 8$ and we sampled $\alpha$ and $\kappa$ uniformly from respective intervals $[-10, 10]$ and $[0, 10]$,  $(\alpha, \kappa) \sim \mathcal{U}([-10, 10]\times [0, 10])$, to inspect the region containing the phase boundaries. In this manner we obtained 6000 $(\alpha, \kappa)$ pairs and their respective configurations $\bX$. The generated data are available from Zenodo~\cite{zenodo}.

\subsection{Descriptors}\label{sec:descriptor}

Next, in the stage B, we introduce a set of physically motivated descriptors. The descriptors substantially reduce the dimensionality of the problem and provide an interpretable representation that facilitates the identification of distinct phases.

A straightforward descriptor for a periodic system is structure factor defined as
\begin{equation}
    {\bm S}_{\bm F}(\bq)=\frac{1}{N^2}\sum_{\bm r} {\bm F}_{\bm r}\, e^{i\bq \cdot \br},
\label{Eq:}
\end{equation}  
where ${\bm F}$ is the relevant field ${\bm P}$ or ${\bm A}$, 
$\bq$ is a wave vector, $N^2$ is the number of the lattice sites, 
and $\bm r$ is a real-space position on the lattice. 

In order to quantify relationship between the polarization field $\bm P$ and the hidden field $\bm A$, we defined correlations as
\begin{equation}
C(\bq) = \frac{\bS_P(\bq)\cdot \bS_A(\bq)}{||\bS_P(\bq)|| \, || \bS_A(\bq)||}
\end{equation}

For the intensities and correlations we assume the 
following set of wave vectors
\begin{align}
    \bQ^\Gamma = (0, 0)\,, \bQ^{X1} = (\pi, 0)\,,  \bQ^M = (\pi, \pi)\,,  \nonumber\\
    \bQ^{X2} = (\pi/2, 0)\,,  \bQ^\Sigma = (\pi/2, \pi/2)\,. \label{wavevectors}
\end{align}

To account for the fourfold rotational symmetry of the PZO potential, the structure factors and correlations were taken
as a mean over all $\pi/2$ rotations of the wave vectors given above
\begin{align}
    \overline{S_X}(\bq) = \frac{1}{4}\sum_{n=1}^4   S_X\!\left(R\left(n\frac{\pi}{2}\right)\bq\right), \\
    \overline{C}(\bq) = \frac{1}{4}\sum_{n=1}^4   C\!\left(R\left(n\frac{\pi}{2}\right)\bq\right),
\end{align}
where $R(\theta)$ is the 2D rotation matrix through angle $\theta$ about origin. We then consider real-valued intensities $\its{\bm F}({\bm Q}) = ||\str{\bm F}(\bQ)||^2$ of the complex-valued structure factors. We define a descriptor vector $\op \in\mathbb{R}^{15}$ for each configuration $\bX$ as
\begin{align}
\op = \big(\its{\bp}(\bm Q^{\Gamma}),\its{\bp}(\bQ^{X_1}), \its{\bp}(\bQ^{M}), \its{\bp}(\bQ^{X_2}), \its{\bp}(\bQ^{\Sigma}), \nonumber \\ \nonumber  \its{\bA}(\bQ^{\Gamma}),\its{\bA}(\bQ^{X_1}), \its{\bA}(\bQ^{M}), \its{\bA}(\bQ^{X_2}), \its{\bA}(\bQ^{\Sigma}), \\ 
\ccor(\bQ^{\Gamma}), \ccor(\bQ^{X_1}), \ccor(\bQ^{M}), \ccor(\bQ^{X_2}), \ccor(\bQ^{\Sigma}) \big).
\end{align}

Using the descriptor vectors $\op$, we constructed preliminary phase labels,
which served both as reference labels for evaluation and as proxy labels for
contrastive learning. The labels were obtained by clustering the descriptor vectors with
DBSCAN~\cite{ester_density-based_1996} as implemented in
scikit-learn 1.8.0~\cite{scikit-learn}. DBSCAN is density-based algorithm
that does not require the number of clusters to be specified in advance. Clustering was performed using the
cosine distance, with the neighborhood radius set to $\varepsilon = 0.01$ and other hyperparameters kept default.

\subsection{Representation learning}\label{sec:encoder}

In the main part of the workflow we trained a neural network encoder $f$ to map the $\bX$ configurations to embedding vectors $f(\bX) = \bm z \in \mathbb{R}^D$, where $D \in \mathbb{N}$ is a chosen dimension of the embedding space. We focus on incorporating the physical information (provided by the descriptors) to the information provided by the raw microscopic configurations $\bX = (\bp, \bA)$. 

The encoder was trained using contrastive learning. Contrastive learning aims to learn an embedding space in which similar data points (samples) are mapped to nearby points, while dissimilar samples are pushed apart. In practice, the objective is formulated using pairs or tuples of samples. For a given \emph{anchor} sample, a \emph{positive} sample is one that is expected to be close to the anchor in the embedding space, whereas a \emph{negative} sample is one that should be mapped far from it. 

Depending on how positive and negative relations are constructed, two common settings can be distinguished. 
In a supervised formulation, pairs are explicitly labeled as positive or negative, 
and the network is optimized to pull positive pairs together in the embedding space while pushing negative pairs above a given margin~\cite{Hadsell_2006_DrLIM,Chopra_2005_SimilarityMetric}. In a self-supervised formulation, positive pairs are generated by applying some transformations to the same input (e.g. rotations or cropping), and negatives are provided by other samples  (often within the same mini-batch); a representative example is SimCLR~\cite{chen_simple_2020}.
In the present work, we explore two strategies for contrastive learning. Specifically, we employ a semi-supervised, margin-based contrastive loss with two different strategies: 
(i) we utilize the clustering-based proxy labels from Sec. \ref{sec:descriptor} to construct the positive/negative relations~\cite{Hadsell_2006_DrLIM, Chopra_2005_SimilarityMetric}, and 
(ii) we use distances in descriptors space directly to define \emph{soft} similarity targets (soft contrastive learning)~\cite{thoma_soft_2020}. In both cases, the resulting embeddings are informed by the microscopic fields $\bp$ and $\mathbf{A}$ as well as by their associated order parameters $\op$.

\begin{figure}[h]
\centering

\newcommand{\encTrap}[4]{%
\begin{tikzpicture}[baseline=(t.base)]
  \def\W{#1}
  \def\H{#2}
  \def\a{#3}
  \coordinate (NW) at (0,\H);
  \coordinate (NE) at (\W,\H);
  \coordinate (SW) at (0,0);
  \coordinate (SE) at (\W,0);
  \coordinate (BL) at ($ (SW)!\a!(SE) $);
  \coordinate (BR) at ($ (SE)!\a!(SW) $);

  \filldraw[draw=TealBlue!80!Black, fill=TealBlue!20, rounded corners=2pt]
    (NW) -- (NE) -- (BR) -- (BL) -- cycle ;

  \node[scale=1.1] (t) at (0.5*\W,0.8*\H) {#4};
\end{tikzpicture}%
}

\tikzset{
    box/.style={
        draw=Black!80, thin, scale=1.1, rounded corners=2pt, rectangle, 
        minimum width=0.8cm, minimum height=0.8cm, inner sep=2pt, align=center},
    input/.style={box, draw=Goldenrod!80!Black, fill=Goldenrod!15},
    output/.style={box, draw=Melon!80!Black, fill=Melon!15},
    dist/.style={box, draw=Lavender!80!Black, fill=Lavender!15, minimum width=2.1cm},
    loss/.style={box, draw=YellowGreen!80!Black, fill=YellowGreen!15, minimum width=1.2cm},
    arrow/.style={-latex, thick, Black!80},
    shared/.style={latex-latex, thick, Black!80, dashed},
    encNode/.style={
        inner sep=0pt, outer sep=0pt, minimum width=0.4cm, minimum height=0.8cm,
        draw=none, fill=none},
    inpbox/.style={
        draw=NavyBlue!80!Black, thick, dashed, text=Black
    },
}

\begin{tikzpicture}[node distance=0.5cm]
\node[input] (x1) {${\bX}_1$};
\node[encNode, below=8mm of x1] (l1)  {\encTrap{1.6cm}{1.8cm}{0.30}{Enc.}} ;
\node[output, below=of l1] (f1) {${\bm z}_1$};

\node[input, right=3cm of x1] (x2) {${\bX}_2$};
\node[encNode, below=8mm of x2] (r1)  {\encTrap{1.6cm}{1.8cm}{0.30}{Enc.}} ;
\node[output, below=of r1] (f2) {${\bm z}_2$};

\node[dist] (dist) at 
    ($(f1)!0.5!(f2) - (0,1.5cm)$) {$\mathrm{dist}({\bm z}_1, {\bm z}_2)$} ;
\node[loss, below=of dist] (loss) {$\lossctr$};

\node[inpbox, inner sep=0.3cm, fit=(x1)(x2)] (inputs) {inputs} ;

\draw[arrow] (x1) -- (l1.north);
\draw[arrow] (l1.south) -- (f1);

\draw[arrow] (x2) -- (r1.north);
\draw[arrow] (r1.south) -- (f2);

\draw[arrow] (f1) -- (dist);
\draw[arrow] (f2) -- (dist);

\draw[arrow] (dist) -- (loss);

\draw[shared] (l1.east) -- (r1.west);

\node[box, align=center, draw=NavyBlue!80!Black] (mid) at ($(inputs.east) + (1cm,-3cm)$) {$\lambda$};

\draw[arrow, draw=NavyBlue!80!Black] 
  -- (inputs.east)
  -| (mid.north) ;
\draw[arrow, draw=NavyBlue!80!Black]
  -- (mid.south) 
  |- (loss.east) ;

\node at ($(l1)!0.5!(r1)$) [yshift=0cm, align=center] {shared\\ weights};
\end{tikzpicture}
\caption{Siamese neural network architecture used for contrastive learning.
${\bm X}_1 \equiv ({\bm P}_1, {\bm A}_1)$ and ${\bm X}_2 \equiv ({\bm P}_2, {\bm A}_2)$ are two independent inputs, and $\lambda \in \{0, 1\}$ carries
information whether they belong to the same cluster or not. 
\emph{Enc.} marks two identical encoder neural networks sharing their internal parameters, which encode the inputs into the
embedding space vectors, ${\bm z}_1$ and ${\bm z}_2$, respectively.
Finally, the contrastive loss $\mathcal{L}_\mathrm{contr}$ 
is calculated from the distance between the encoded vectors, 
$\mathrm{dist}({\bm z}_1, {\bm z}_2)$ and the values of $\lambda$.}
\label{siamese}
\end{figure}
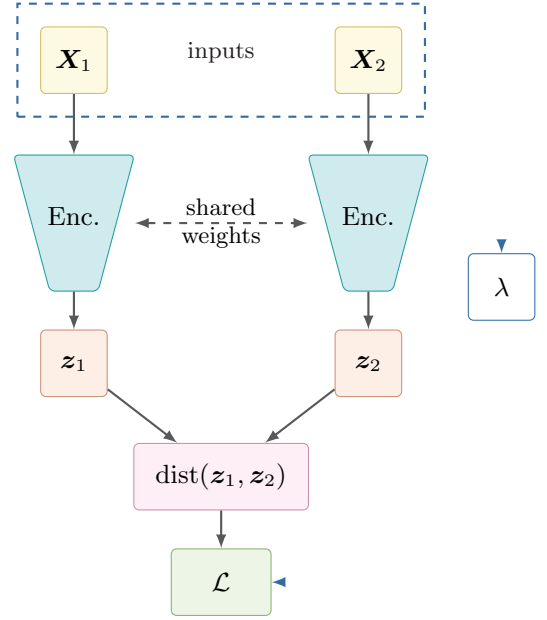

\subsubsection{Contrastive learning}
\label{sec:contrastive}

In the (classical) contrastive learning, the batches are constituted of pairs, where one configuration is the anchor and the other configuration is either negative or positive. To find the encoder weights, a loss function depending on the distance between the embeddings $z$ and on a label variable $\lambda$ is minimized. The variable $\lambda$ carries the information about the type of pair being evaluated and its value is obtained from the inputs so that for two configurations $\bX_i$ and $\bX_j$,
    \begin{equation}
    \lambda_{ij} = \begin{cases} 
          1 & \text{$\bm X_i$ and $\bm X_j$ are positive} \\
          0 & \text{otherwise}.
          \end{cases}
\end{equation}
The full loss functions sums over all the combinations in a batch and reads
{\small
\begin{equation}
\lossctr = \frac{1}{2} \sum_{i \neq j} \left[ \lambda_{ij}\, d_{ij}^2 + (1 - \lambda_{ij})\, \left(\max\left\{0, m - d_{ij} \right\} \right)^2 \right]\,,\label{discrete_loss}
\end{equation}}
where $d_{ij}$ is a value of a distance function for the embedding vectors $d_{ij} := \mathrm{dist}(f(\bm X_i), f(\bm X_j))$ and $m \in \mathbb{R}$. The parameter $m$ in \eqref{discrete_loss} controls the minimal distance between the negatively labeled pairs of embeddings and we set $m = 1$. We label the classical contrastive learning embeddings as $\bm v$. We chose a simple euclidean distance on the embedding space $\mathrm{dist}({\bm v}_i, {\bm  v}_j) = || {\bm v}_i - {\bm v}_j||$. The encoder was trained using a Siamese architecture depicted in Figure~\ref{siamese}, where each pass is computed in pairs and the encoder in both branches share its weights. The details of the encoder architecture are given in Appendix \ref{app:encoder}. The full implementation, including all hyperparameters and training schedules, is available in Ref.~\cite{repo}.

For a contrastive learning model to perform well, especially in a self-supervised setting, the training pairs must be chosen carefully, since the sampling strategy can be just as important as selecting an appropriate loss function~\cite{Wu_2017_ICCV}. In our case, the type of contrast is determined by the preliminary phase labels obtained from clustering on the space of descriptors outlined in Sec.~\ref{sec:descriptor}: configurations are taken as a positive pair whenever they belong to the same cluster, and as a negative pair otherwise. Since we already possess these labels, we simply consider batches containing equal numbers of positive and negative pairs, without imposing any additional constraint on which cluster the negative samples are drawn from. This simple strategy is possible mainly due to the simplicity of the dataset and the supervised nature of the task. Furthermore, we have applied data augmentation, implemented as random in-plane rotations of the contrast configurations by an angle $k\pi/2$ with $k \in \{1,2,3\}$.

\subsubsection{Soft contrastive learning}
In the case of the soft contrastive learning approach, to compare  $\bX_i$ and $\bX_j$, instead of a hard label $\lambda_{ij} \in \{0,1 \}$ indicating whether the two samples belong to the same class or not, a soft label $\sl \in [0,1]$ is assigned. The soft label $\sl$ expresses continuous degree of similarity between the two samples. For given configurations $\bX_i$, $\bX_j$, we calculated the soft label $\sl$ from their respective descriptor vectors $\op_i$, $\op_j$ as the cosine similarity,
\begin{equation}
    \sl = \frac{\op_i^T\cdot \op_j  }{||\op_i||\, || \op_j|| \, }\,.
\end{equation}
Furthermore, the values of $\sl$ were min-max scaled with respect to the whole dataset, so that the values of $\sl$ span the full $[0, 1]$  .

\begin{figure}
    \centering
    \includegraphics[width=0.7\linewidth]{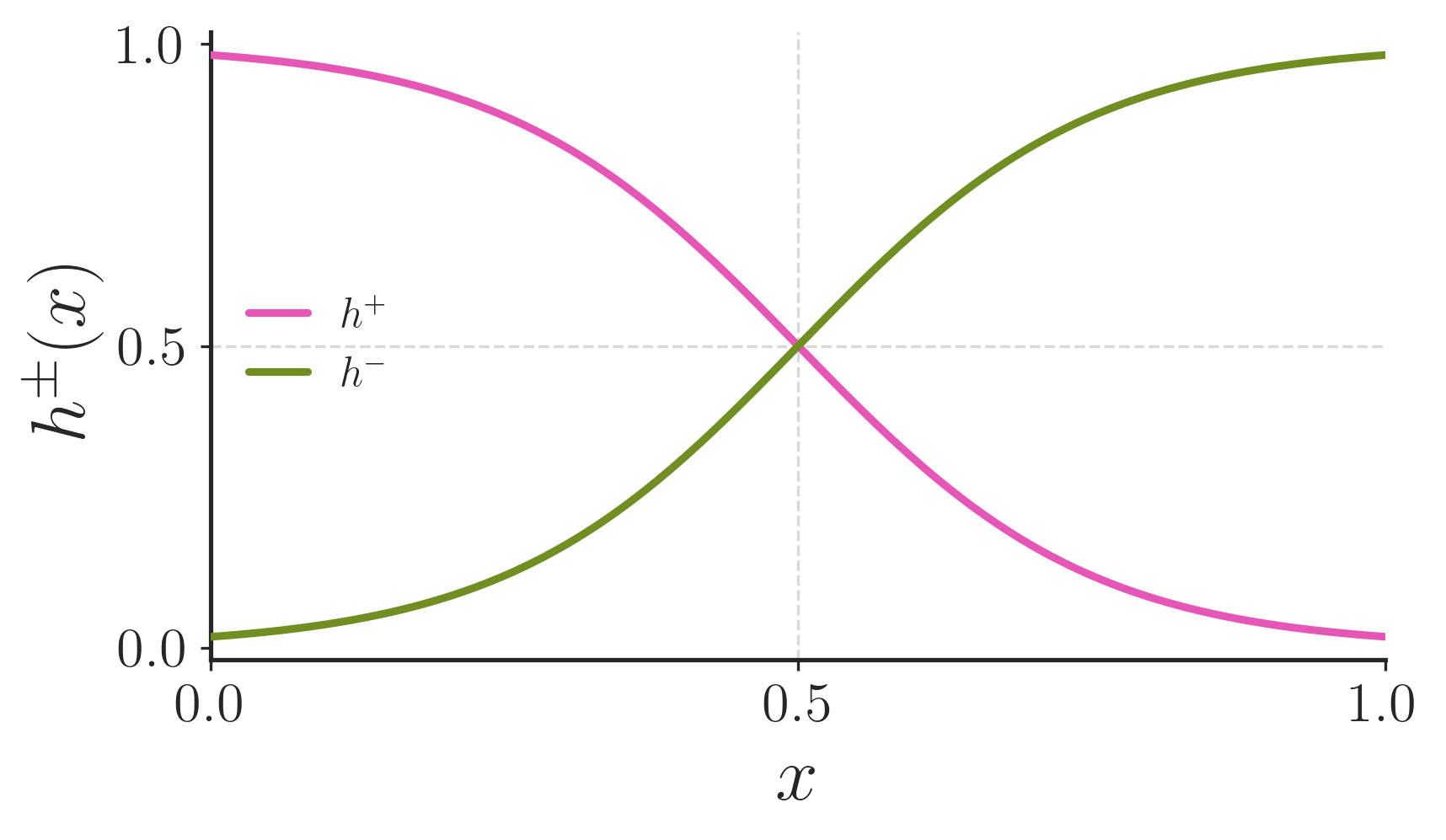}
    \caption{Graph of functions $h^\pm$ on the interval $[0, 1]$ for parameters $a= 8$ and $b = 0.5$. The functions $h^\pm$ and their parameters control how the $\sl$ similarity influences the final loss $\losssoft$ values.}
    \label{fig:functions}
\end{figure}

We then considered a generalized version of the contrastive loss \eqref{discrete_loss}:
{\small
\begin{equation}
\losssoft = \frac{1}{2} \sum_{i \neq j} \left[ h^+(\sl)\, d_{ij}^2 + h^-(\sl)\, \left(\max\left\{0, m - d_{ij} \right\} \right)^2 \right]\,,\label{cont_loss}
\end{equation}}
where $h^+$ is a decreasing real-valued function and $h^-$ is an increasing real-valued function, such that $h^+ + h^- = 1$. Functions $h^\pm$ and their parameters allow for tuning how the value of the label $s_{ij}$ influences the final loss, effectively controlling the decision boundary between positive and negative samples. The $m \in \mathbb{R}$ parameter plays the same role as in the contrastive loss given in \eqref{discrete_loss} and was set as $m=1$. We denote the embeddings from soft contrastive learning as $\bm w$ and similarly choose $d_{ij} = || \bm w_i - \bm w_j||$. For functions $h^\pm$, we used the sigmoid functions
\begin{equation}
    h^{\pm}(s_{ij}) = \frac{1}{1 + \exp\left[\mp a(s_{ij} - b)\right]},
\end{equation}
where $a, b \in \mathbb{R}^+_0$. We set $a = 8$ and $b = 0.5$, where the $b$ choice reflects corresponds to the fact that the soft label $s_{ij}$ takes value in the interval $[0,1]$. The functions are depicted in the Figure \ref{fig:functions}.
 The soft contrastive learning approach therefore allows us to skip the unsupervised preliminary label generation and directly obtain the vectorial representation of the data. 

As for the encoder architecture, it was identical to the contrastive learning encoder and the details are specified in Appendix~\ref{app:encoder}. Training relied on a shared-weight encoder. Shared-weight encoder is a generalization of the Siamese encoder in which a single forward pass is applied to the entire batch rather than to a pair of inputs. The implementation and sample training is available at Ref.~\cite{repo}.

\subsection{Phase diagram construction}\label{sec:phaseid}

In the final part of the workflow, we use the trained encoder (Sec.~\ref{sec:contrastive}) to identify distinct phases in the training dataset.

We first obtained the embeddings $\bm z$ for all MC configurations $\bX$.
Clusters in the embedding space were identified using HDBSCAN, a hierarchical
extension of the DBSCAN algorithm~\cite{campello_density-based_2013}, as implemented in the scikit-learn library, with
the Euclidean distance as the metric, $\texttt{min\_cluster\_size} = 30$, $\texttt{cluster\_selection\_epsilon} = 0.1$ (classical) and $\texttt{cluster\_selection\_epsilon} = 0.02$ (soft contrastive). Other hyperparameters of the HDBSCAN were set to default. For each cluster, we selected as representatives the $n = 3$ configurations whose embeddings have the smallest average distance to all other members of the same cluster.

We then sampled the region $[-10, 10] \times [0, 10]$ on uniform $200\times100$ grid, which is considerably denser than the grid used for the MC simulated annealing. At each grid point, we ran a Broyden--Fletcher--Goldfarb--Shanno algorithm (BFGS) optimization, as implemented in the SciPy library \cite{2020SciPy-NMeth}, starting from every cluster representative found. We identified the ground state as the relaxed configuration with the lowest energy. Testing all cluster representatives at every point is affordable due to the low computational cost of BFGS.

This procedure yields a dense grid in the $(\alpha, \kappa)$ space with a well-relaxed configuration $\bX$ at every point. To obtain the phase diagram, we mapped the configurations into the embedding space using the encoder trained on the MC configurations. The embeddings were then clustered with HDBSCAN to obtain the phase diagram,
using the same hyperparameters as for the MC embeddings except for
\texttt{min\_cluster\_size}~$= 50$ and \texttt{cluster\_selection\_epsilon}~$= 0.015$.

\section{Results}\label{Sec:Results}
 
In this section, we first examine the results obtained from the encoders in the case of the Monte--Carlo generated configurations for both the classical and the soft contrastive learning. Next, we show how the method handled the gradient-optimized ground-state configurations, leading to the final phase diagram. We briefly recall the notation. Embedding coordinates are denoted by lowercase symbols: $\bm v=(v_1,v_2)$ (classical contrastive), and $\bm w=(w_1,w_2,w_3)$ (soft contrastive). Cluster labels use uppercase identifiers $\mathrm{V}i$, and $\mathrm{W}i$ for the respective methods and $\mathrm{U}i$ for the preliminary clusters obtained by direct clustering of the $\op$ vectors.

\begin{figure}[h]
    \centering
    \includegraphics[width=\linewidth]{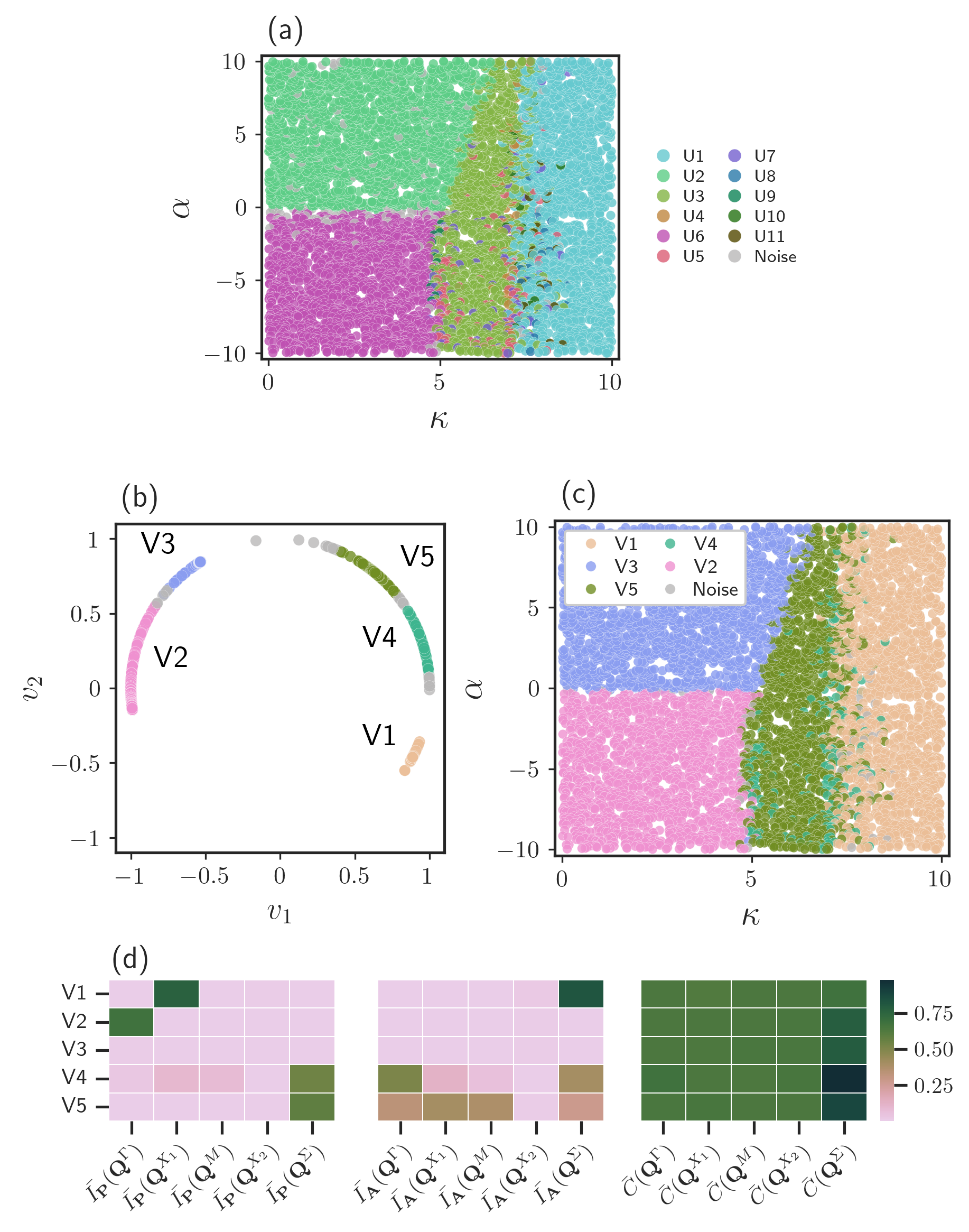}
\caption{Classical contrastive learning results for configurations obtained from MC simulated annealing. (a) Preliminary clusters found by clustering directly on the space of descriptor vectors $\op$ shown across the parameters plane $(\alpha, \kappa)$. (b) The embeddings and clusters obtained from the contrastive learning encoder. (c) The cluster assignments of embeddings from (b) shown across the $(\alpha, \kappa)$ plane. (d) Mean values of the descriptors for found clusters. Prior to the mean calculation, the individual descriptor components $\chi_i$ were scaled so that $\chi_i \in [0,1] $.} 
\label{fig:contrastive_result}
\end{figure}

The results for the classical contrastive learning are given in Fig.~\ref{fig:contrastive_result}. Panel~(a) shows the preliminary cluster labels plotted against the model parameters $\alpha$ and $\kappa$, giving overview of the phase diagram. Panel~(b) shows the embeddings obtained from the contrastive learning encoder, while  panel~(c) plots the clusters from panel~(b) against $(\alpha, \kappa)$, giving comparison to the preliminary labels given in (a). Finally, panel~(d) reports the mean values of the individual (normalized) descriptor vector components $\chi_i$ inside each cluster.

Compared to preliminary labels, the contrastive learning encoder was able to reduce number of labels from 11 to 5, however few points were assigned to a noise class. The noise class groups samples that did not satisfy the criteria to be assigned to any cluster with respect to the hyperparameters of the used clustering algorithm. From the mean values of the order parameters shown in Fig.~\ref{fig:contrastive_result}(d) and by inspection of the configurations, it is evident that the V2 cluster is composed of FE configurations and the V3 cluster consists of PE configurations. The remaining clusters V1, V4 and V5 feature different forms of stable or metastable AFE configurations. We note, that the contrastive clusters V1, V2 and V3 were mostly identical to the respective preliminary clusters U1, U6 and U2. The difference was a multitude of smaller clusters of perturbed configurations inside the AFE region, which were redistributed across 3 contrastive learning AFE clusters (V1, V4 and V5). Furthermore, Fig.~\ref{fig:contrastive_result}(b) reveals geometric structure of the embedding space: the PE and FE phases occupy adjacent regions of the space, while the the AFE-related embeddings are located in different arcs of the circle.

\begin{figure}
    \centering
    \includegraphics[width=1.0\linewidth]{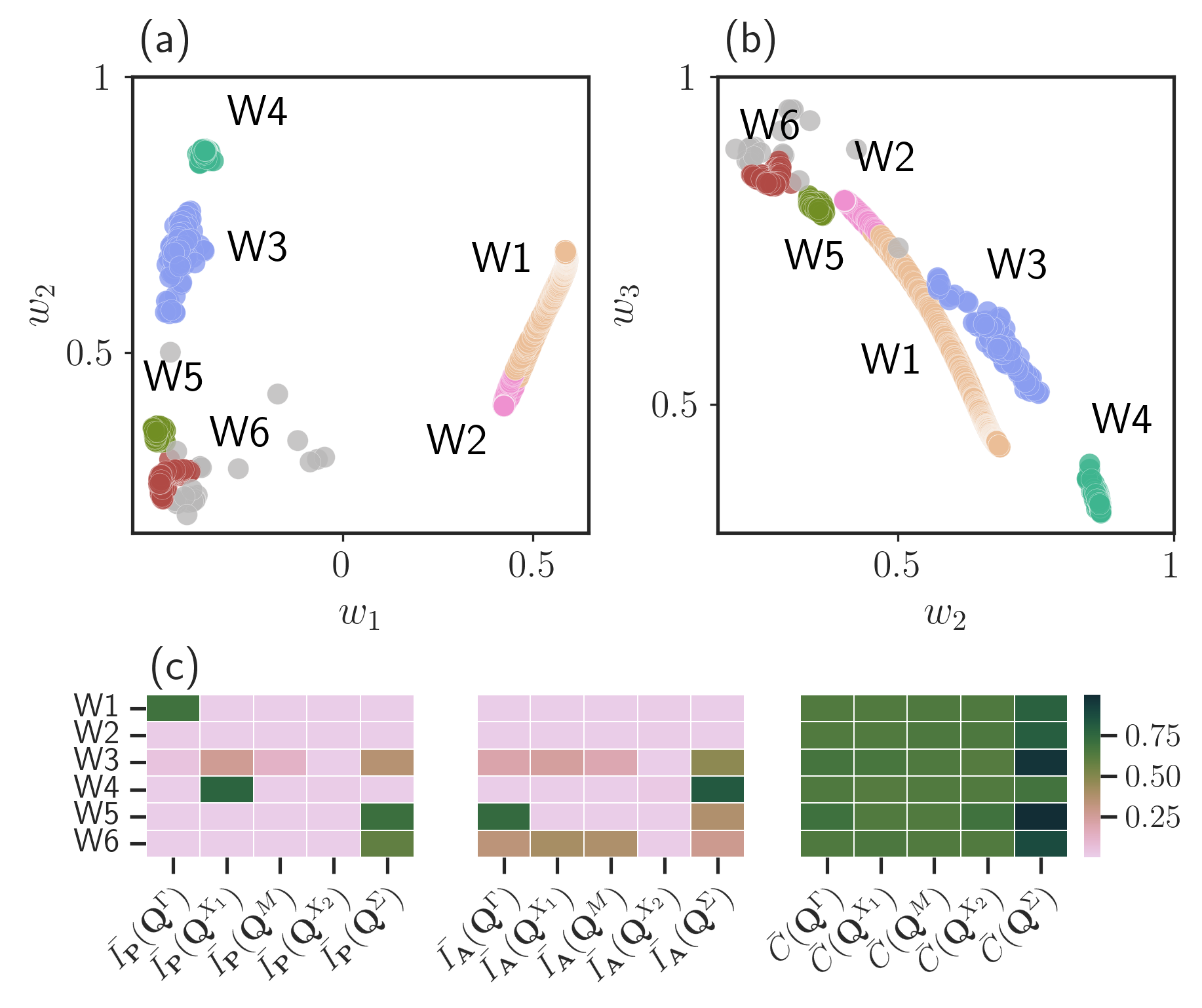}
    \caption{Soft contrastive learning results for configurations obtained from simulated annealing (a-b)  Three-dimensional embedding obtained from the soft contrastive learning encoder, shown as projections onto the $(w_1, w_2)$ and $(w_2, w_3)$ planes and annotated with cluster assignment. (c) Mean values of the descriptors for found clusters. Prior to the mean calculation, the individual descriptors $\op_i$ were scaled so that $\op_i \in [0,1] $.}
    \label{fig:sft_ctr_result}
\end{figure}

The soft contrastive learning algorithm yielded similar results to the classical contrastive learning and these results are depicted in Fig.~\ref{fig:sft_ctr_result}. We chose embedding dimension $D=3$ as is discussed in Sec.~\ref{Sec:Discussion} and more in detail in Appendix~\ref{app:collapse}. The embeddings are shown in panels (a) and (b) as projections to $(w_1, w_2)$ and $(w_2, w_3)$ planes with found cluster labels.  Panel (c) depicts heatmap with mean values of the descriptors $\op_i$ for the individual clusters, using the same scaling of $\op_i$ as in Fig.~\ref{fig:contrastive_result}(d). The results are consistent with the clusters obtained in the classical contrastive learning method, apart from one additional cluster being found in the soft contrastive variant. Specifically, clusters W1 and W2 correspond to PE and FE phases respectively and the rest of the clusters group various forms of AFE configurations. The dependency on the parameters $\kappa$ and $\alpha$ is similar as shown for the case of classical contrastive learning given in Fig.~\ref{fig:contrastive_result}(c), and is therefore omitted. One important difference between the results is the smooth variation of the embedding vectors between PE and FE regions (clusters W1 and W2), which can be attributed to the continuous variation in relevant structure factor $I_P(Q^{\Gamma})$ between these two regions. 

\begin{figure}
    \centering
    \includegraphics[width=1.0\linewidth]{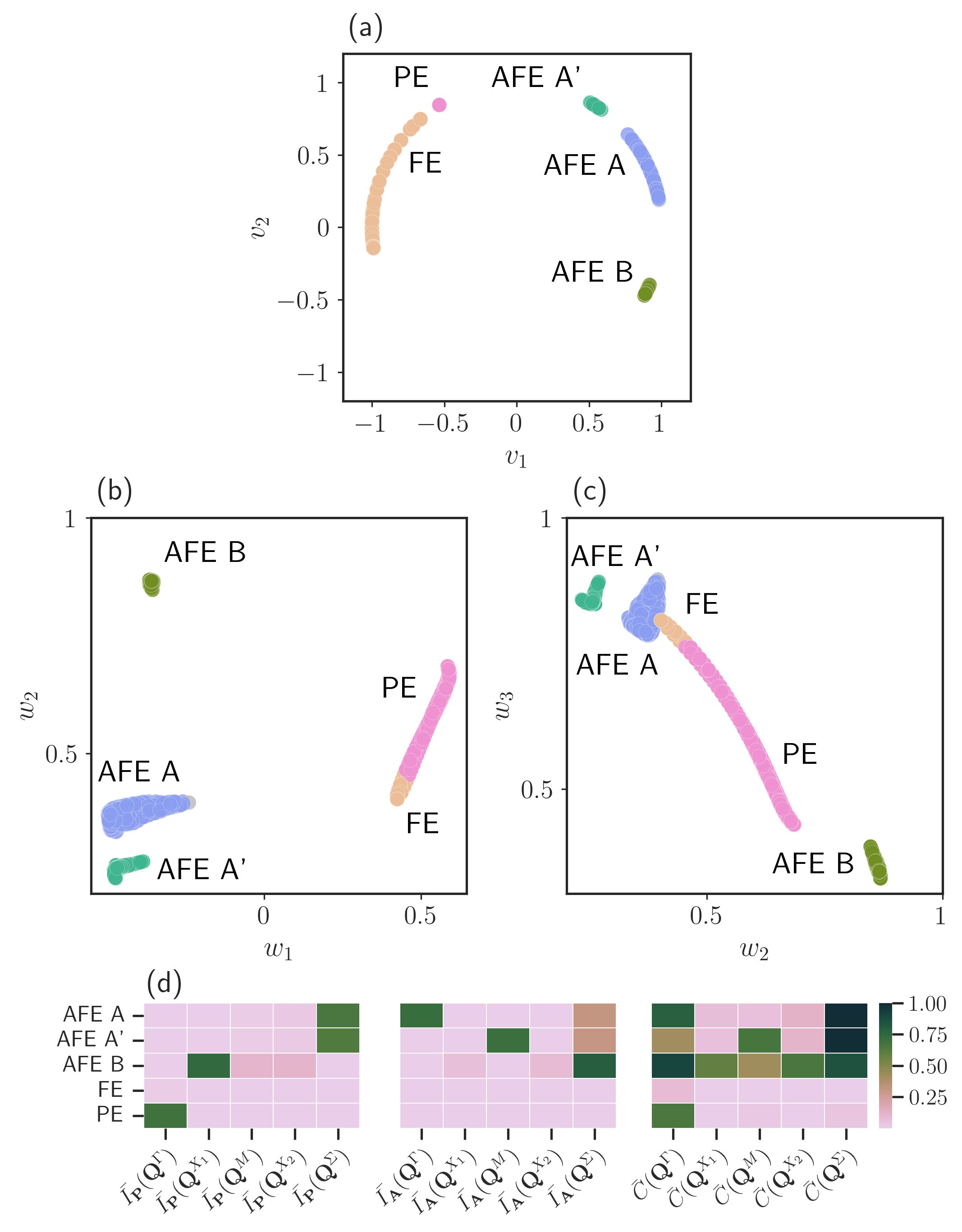}
    \caption{Classical and soft label contrastive learning results for BFGS-optimized ground-state configurations. (a) Embeddings and clusters obtained for classical contrastive encoder. (b-c) Embeddings and clusters obtained from the soft contrastive learning encoder, shown as projections onto the $(w_1, w_2)$ and $(w_2, w_3)$ planes. (c) Mean values of the descriptors for found clusters. Prior to the mean calculation, the individual descriptors $\op_i$ were scaled so that $\op_i \in [0,1] $.}
    \label{fig:final_embeddings}
\end{figure}

\begin{figure}
    \centering
    \includegraphics[width=1.0\linewidth]{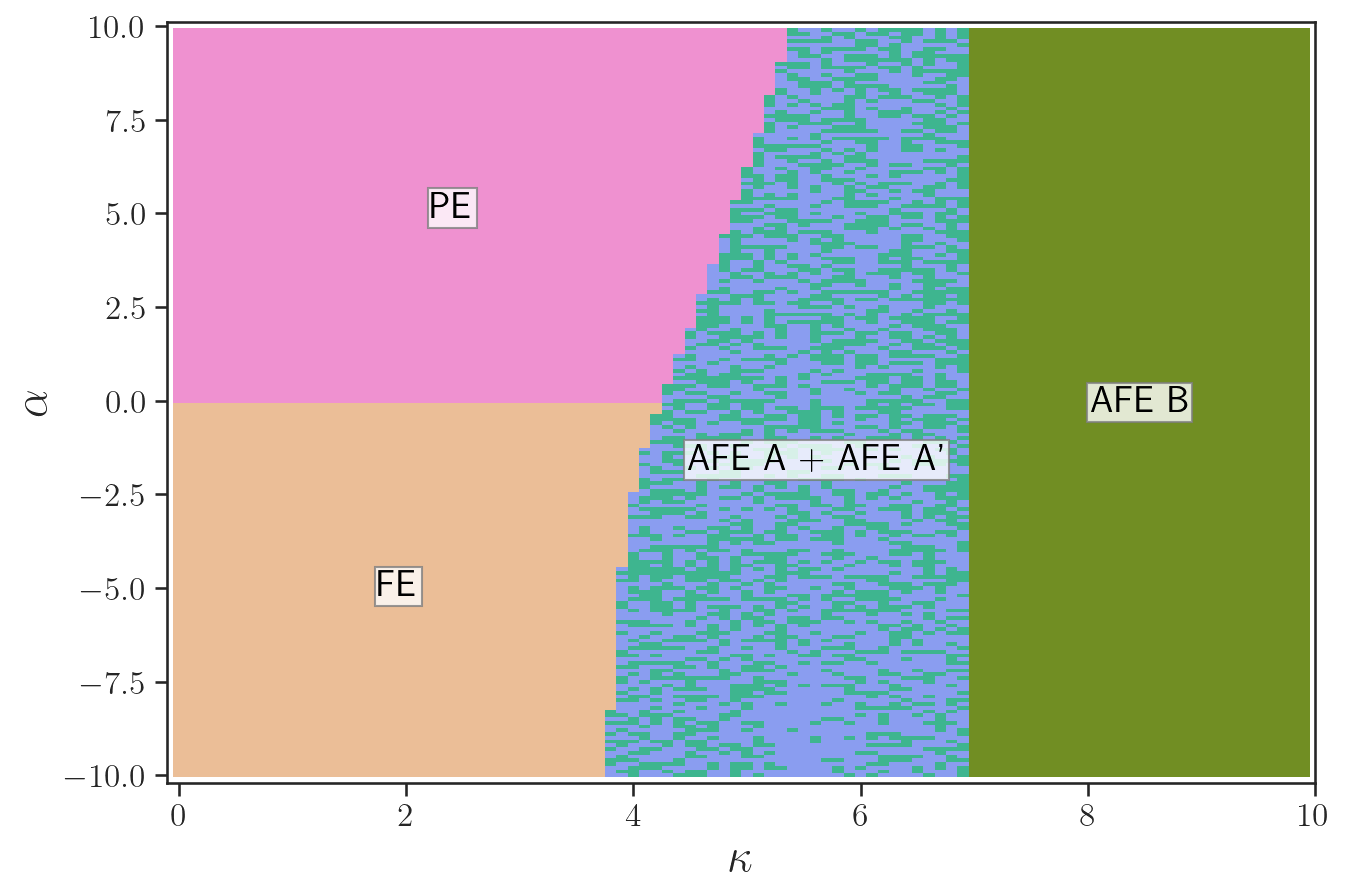}
    \caption{Phase diagram in the $(\alpha,\kappa)$ parameter plane obtained from the BFGS-optimized ground-state configurations, annotated by found phase (cluster) label.}
    \label{fig:phase_diag}
\end{figure}

\begin{figure}
    \centering
    \includegraphics[width=0.9\linewidth]{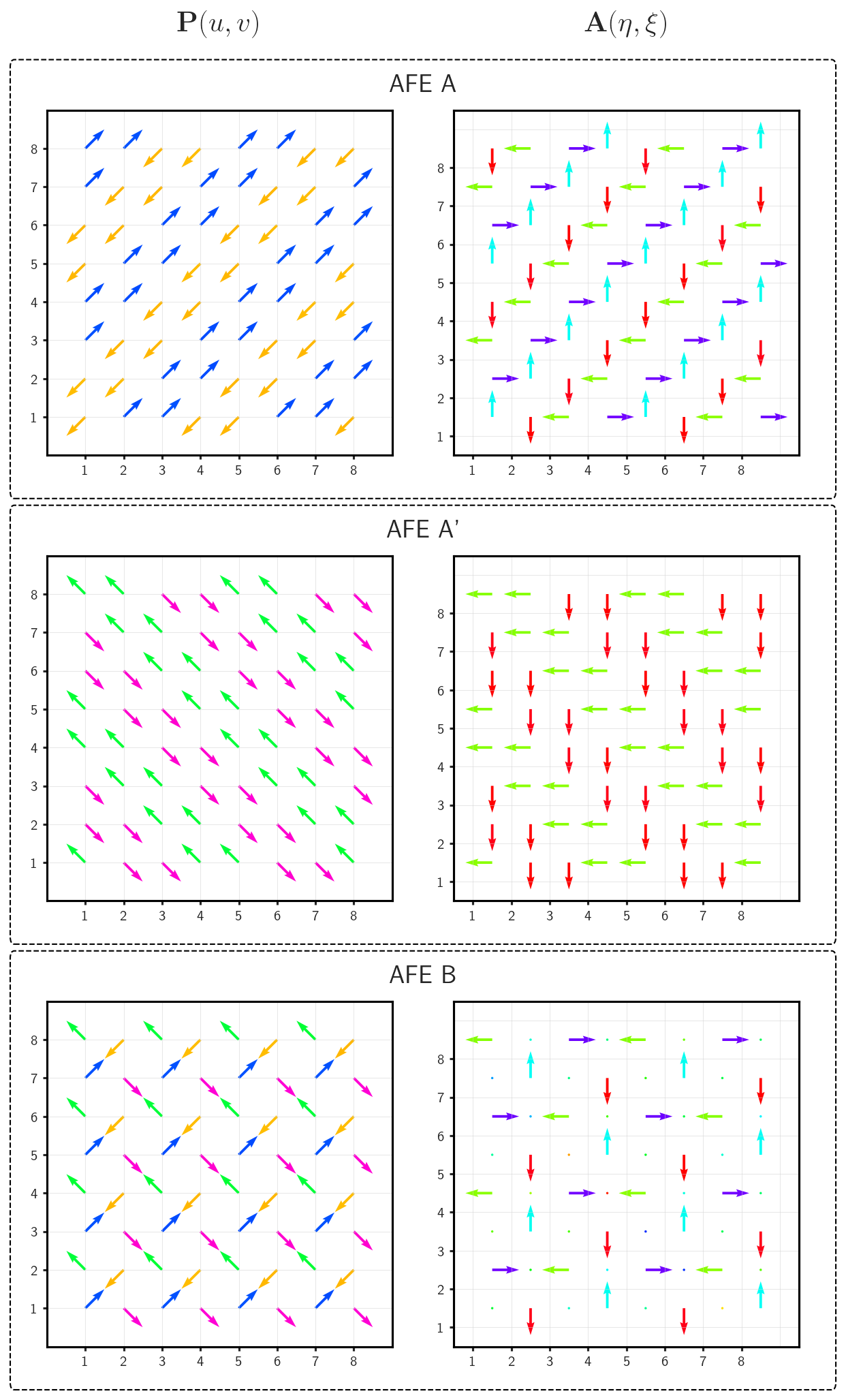}
    \caption{Representatives of the relaxed AFE~A. AFE~A' and AFE~B phase configurations. The left column shows the polarization field $\bp(u, v)$ and the right column the hidden field $\bA(\eta, \xi)$. The arrows represent field directions only, as their lengths are not drawn to scale.}
    \label{fig:Configs}
\end{figure}

As for the deterministic optimization, we observed that the important clusters representing the PE, FE, AFE A and AFE B phases were found and preserved by both encoders. Initial configurations from both methods therefore lead to the correct ground state configurations after the BFGS optimization. The embeddings of the ground states from both methods are given in Fig.~\ref{fig:final_embeddings} and the finalized phase diagram is given in Fig.~\ref{fig:phase_diag}. The results for both methods are equivalent, except for noise points at the region boundaries in the soft-contrastive case, Fig.~\ref{fig:final_embeddings}(a-b). In agreement with the results obtained from the MC configurations, we observed the four aforementioned phases. Furthermore we found that the model allowed for two different configurations in the AFE A domain which we call AFE A and AFE A'. Chosen representatives of the 3 different AFE configurations are visualized in Fig.~\ref{fig:Configs}. 

\section{Discussion}\label{Sec:Discussion}

In Sec. \ref{Sec:Results} we have presented all parts of our workflow, described in
Fig.~\ref{fig:diagram}, which results in construction of the final phase diagram shown in Fig.~\ref{fig:phase_diag}. Let us now analyze the crucial parts of the pipeline.

\subsection{Use of descriptors}

First, a supporting physical information enters the workflow through the training signal in a form of a vector of physics-motivated descriptors $\op$ defined in Sec.~\ref{sec:descriptor}, consisting of structure-factor intensities and correlations at a small set of high-symmetry wave 
vectors, symmetrized over the point group of the lattice. These descriptors are 
used solely to define the notion of similarity between two configurations: in the 
classical branch they generate the positive and negative pairs, and in the 
soft-label branch they enter directly as a continuous target similarity. The 
encoder therefore retains access to the full configuration, while the geometry it 
is asked to reproduce in the embedding space is the one expressed by $\op$. Since 
the descriptors are invariant under lattice translations and symmetrized over its 
point group, the target similarity inherits these invariances, and the encoder is 
driven towards a representation that respects them.

For simplicity, the descriptor set used here is not complete. Two states whose 
modulations are carried by wave vectors outside the chosen set would be assigned 
the same target similarity as a pair of identical states, and the contrastive 
objective would map them onto the same region of the embedding space even though 
the underlying configurations differ. Here this is not a limitation, since the 
wave vectors are dictated by the symmetry and the interaction range of the 
potential. When the underlying model is not known in advance, the same workflow 
should be run with a denser sampling of wave vectors, or with descriptors learned 
directly from the configurations under an imposed equivariance 
constraint~\cite{cohen_2016:ICML}.

Second, to construct positive and negative pairs for the classical
contrastive method, we used DBSCAN clustering in the space of descriptors making use of cosine metrics. The resulting clusters for the MC configurations are shown in
Fig.~\ref{fig:contrastive_result}(a). The four dominant ones, U1, U2, U3 and U6,
correspond to the AFE~B, PE, AFE~A and FE phases, respectively, so that the main
phases of the model are already recovered at this stage. That a coarse phase
diagram can be retrieved from the $\op$ vectors alone indicates that the
descriptors already capture the distinctions between the main phases.

This intermediate clustering is, however, not the final result. It 
partitions the configurations into considerably more clusters than the model has 
distinct states, so that the correspondence between clusters and phases has to be 
established by hand, and assigning a new configuration requires evaluating the 
full descriptor vector and matching it against the existing partition. More 
fundamentally, the partition rests on a fixed, prescribed metric in $\op$, in 
which every descriptor component contributes on an equal footing regardless of 
whether it is relevant to the ordering. This is acceptable here, where the wave 
vectors entering $\op$ were chosen with the symmetry and the interaction range of 
this particular potential in mind and nearly every component is informative, but 
it does not carry over to a general setting: with a larger and predominantly 
uninformative descriptor set, the distance between two configurations would be 
governed by components unrelated to the ordering. The contrastive step provides
representation in which the relative importance of the descriptors is determined 
by the data rather than imposed.

\subsection{Contrastive learning methods}

Third, we used the contrastive learning to create embeddings that allow us to identify distinct phases in our data. The classical contrastive encoder trained stably and produced comparable results across independent runs of the optimization, and the HDBSCAN clusters obtained from its embeddings were stable over a wide range of hyperparameters. 
We also observed that data augmentation, based on symmetry operations applied to training dataset, did not improve the results. We attribute this robustness primarily to the simplicity of the input data, the choice of descriptors, and the effectively supervised nature of the contrastive task, in which the proxy labels are fixed before training.

All five states were already resolved in a two-dimensional embedding space. For $D = 3$ we observed a dimensional collapse: the embedding vectors clearly spanned a two-dimensional hyperplane and the result was equivalent to the $D = 2$ case. In self-supervised contrastive learning this effect is usually 
attributed to implicit regularization or to strong augmentation \cite{Li_2022_DimensionalCollapse}. 
Even in the absence of collapse, however, the higher embedding dimension did not carry any additional information about the system. We refer to Appendix~\ref{app:collapse} for further analysis of the descriptors for varying values of $D$.

In contrast to the classical variant, the soft-contrastive method requires no clustering prior to training: the positive and negative pairs are replaced by a graded target similarity computed directly from the descriptor vectors, so that the clustering step in the descriptor space is dropped entirely. 
Clustering is still performed after training, in the learned embedding space, in order to identify the final phase regions.

Note, at the FE--PE boundary the configurations were mapped onto an arc terminating in the PE points with vanishing $I_P(Q^\Gamma)$, as seen in 
Fig.~\ref{fig:final_embeddings}(b-c); the separation of the two regions was more subtle than in the classical case and required careful tuning of the HDBSCAN hyperparameters. Moreover, with suitably tuned HDBSCAN parameters all four regions were always
identified, but AFE~A and AFE~A' were separated more consistently in the final
phase diagram for $D \geq 3$ than for $D = 2$. Furthermore, we found that the soft contrastive learning encoder was more likely to utilize the third dimension for $D\geq3$ than the classical contrastive learning encoder.

Overall, both approaches successfully recovered the phase diagram. Classical contrastive learning refined the proxy-label information and reduced the number of clusters, but consequently remained dependent on the quality of the proxy-label construction step. The soft-label variant removed that step altogether by incorporating the structure-factor relationships directly, and reflected the more nuanced distinctions between the PE/FE and AFE A/AFE A' regions, at the cost of a greater sensitivity to the hyperparameters of the final clustering.

\subsection{Interpretability and identified phases}

Finally, let us analyze the main results of our workflow, which are the identified physical phases.
The use of physically motivated descriptors makes the workflow interpretable at the level at which the phases are identified. The learned embedding itself remains opaque, but every cluster in it can be characterized by its averaged descriptors $\bar{\op}$, which have a direct physical meaning. 
Besides assigning labels to regions of the parameter space, the clustering provides the order parameters that distinguish them, as summarized in Fig.~\ref{fig:final_embeddings}(d).
The interpretation is confined to the chosen set of descriptors: 
in addition to the restriction to a finite set of wave vectors discussed above, the use of intensities discards the relative phase between ${\bm P}$ and ${\bm A}$, so that states differing by a rigid shift of the hidden field with respect to the polarization are not distinguished.

It should be emphasized that the machine-learning step does not itself decide which phase is realized. 
The configurations sampled by Monte Carlo are compressed into a small set of representatives by the contrastive encoder, and these representatives are then used as initial conditions for a gradient-descent relaxation performed independently at each point of the $(\alpha, \kappa)$ grid; the phase is assigned as the relaxed state of lowest energy. 
The selection criterion is thus energetic and deterministic, and the workflow cannot produce a spurious phase. The contrastive step replaces either a manual choice of candidate ansatzes or a large number of random initial conditions by a small set of physically distinct candidates. 
On the other hand, a phase that is never visited during the Monte Carlo sampling cannot enter the set of representatives and will not be tested in the relaxation, so the procedure is exposed to false negatives rather than false positives. A related concern arises for phases that are sampled only rarely, since 
a cluster supported by very few samples may be absorbed into a neighboring one. 
In such cases an analysis of the descriptor vectors prior to training, targeting outliers in the Monte Carlo data, could be used to flag weakly represented states before the contrastive step.

Our workflow identified distinct configurations present in the diagram \ref{fig:phase_diag}. Based on the values of descriptors in each cluster we identified following phases.
The paraelectric (PE) phase is the reference state, with all descriptors vanishing. 
The configuration is given by
\begin{equation}
    r_{m,n} = s_{m,n} = \phi_{m,n} = \psi_{m,n} = 0 
\end{equation}
The ferroelectric (FE) phase is characterized by a nonzero $\overline{I_P}(\bQ^\Gamma)$ 
and no modulation at any finite wave vector. FE configuration satisfies one of the orderings
\begin{subequations}
    \begin{align}
        r_{m,n} &= P\,, s_{m,n} = \phi_{m,n} = \psi_{m,n} = 0\,,\\
        s_{m,n} &= P\,, r_{m,n} = \phi_{m,n} = \psi_{m,n} = 0\,,
    \end{align}
\end{subequations}
where $P$ is the magnitude of local polarization, which can be obtained by minimization of the total potential
(\ref{Eq:potential}) for given $(\alpha, \kappa)$.
The transition of FE towards PE proceeds through a continuous 
decrease of $P$, which is reflected in the arc-like structure of the 
corresponding region in the embedding space shown in Fig.~\ref{fig:final_embeddings}. 

The two states AFE A and AFE A' share the same antipolar pattern of the polarization field, with the dipoles modulated along one of the two diagonals of the lattice and a vanishing total polarization, but differ in the arrangement of the hidden field.
In both states the hidden field carries a component at $\bQ^\Sigma$, commensurate 
with the modulation of the polarization. They differ in the additional component 
that accompanies it: in AFE A the hidden field has a further contribution at 
$\bQ^M$, corresponding to a checkerboard alternation, whereas in AFE A' the 
additional contribution is uniform, at $\bQ^\Gamma$.
The AFE A phase configuration can be expressed as
 \begin{equation}
    \begin{aligned}
        r_{m,n} &= 0\,,\\
        s_{m,n} &= P\, \sqrt{2}\, \cos\left(\frac{\pi}{2}(n-m+3/2)\right)\,,\\
        \phi_{m,n} &= A\, (-1)^{n+m}\,,\\
        \psi_{m,n} &= A\, (-1)^{\left\lfloor (n - m)/2 \right\rfloor}\,,
    \end{aligned}
\end{equation}
where $A$ is the size of the hidden field vectors computable by the minimization of potential (\ref{Eq:potential}).
States AFE A and AFE A' are stable over the same region of the $(\alpha, \kappa)$ plane and are degenerate in energy. This degeneracy arises from a symmetry operation that preserves the energy by flipping the hidden field vectors according to the following transformation,
\begin{equation}
    \begin{aligned}
        \phi_{m,n}^{\prime} &= (-1)^{n+m+1}\phi_{m,n}\,,\\
        \psi_{m,n}^{\prime} &= (-1)^{n+m+1}\psi_{m,n}
    \end{aligned}
\end{equation}
Thus, AFE A and AFE A' represent two realizations of a single phase rather than two distinct regions of the phase diagram. They are nevertheless resolved as separate clusters by the descriptors, which distinguish them through the modulation of the hidden field, a distinction that is by construction invisible to a classification based on energy alone.

The remaining phase, AFE B, has both diagonal components of the polarization present simultaneously, so that neighboring dipoles are mutually perpendicular while the total polarization still vanishes. 
The configuration is given by
\begin{equation*}
    \begin{aligned}
        r_{m,n} &= \frac{P}{2}\, (-(-1)^n + (-1)^m)\,,\\
        s_{m,n} &= \frac{P}{2}\, ((-1)^n + (-1)^m)\,,\\
        \phi_{m,n} &= A\, \cos\left(\frac{\pi}{2}(n + m + 1)\right)\,,\\
        \psi_{m,n} &= A\, \cos\left(\frac{\pi}{2}(n - m - 1)\right)\,.
    \end{aligned}
\end{equation*}

Two of these phases, AFE A and AFE A', correspond to the $\uparrow\uparrow\downarrow\downarrow$ antipolar pattern of the $Pbam$ structure commonly associated with PbZrO$_3$~\cite{sawaguchi_antiferro_1951, corker_pzo_1997,iniguez_pzo_2014,hlinka_2014:PRL}. 
On the other hand, in AFE B the polarization direction rotates from 
site to site, so that the field lines close into a periodic array of loops.
Flux-closure configurations of this kind have been reported in 
PbZrO$_3$ around antivortex cores at ferroelastic twin boundaries 
\cite{Liu_2025_vortices}, although there they occur as a local structure bound to 
a domain wall rather than as a bulk periodic state.

\section{Summary}
\label{Sec:Summary}

In this work, we proposed a data-driven workflow for unsupervised construction of a phase diagram, which we successfully applied to obtain a phase diagram of the suggested 2D PZO model. The workflow utilized contrastive learning, incorporating both the microscopic field configurations and the relationships of physically motivated descriptors. 

The two explored contrastive learning objectives were classical contrastive learning and soft-label contrastive learning. In the classical method, preliminary labels were found from dimensional reduction and clustering of descriptor vectors $\op$ and they were used to generate negative and positive pairs of samples. In the soft contrastive learning, the algorithm used $\op$ vectors directly and the soft label was calculated as their cosine distance.

We have shown that both contrastive learning objectives were able to retrieve structure from the metastable configurations and suggest good initial data for the gradient-descent based optimization algorithm. After final configuration optimization, both obtained encoders provided latent space representations of the ground state configurations that formed a well-structured clusters corresponding to the paraelectric, ferroelectric and few antiferroelectric configurations. Notably, the workflow resolved AFE A and AFE A' as separate clusters although the two states are degenerate in energy and occupy the same region of the $(\alpha, \kappa)$ plane. 

The workflow itself is not tied to the particular potential studied here. It requires only a sampling procedure for the configurations and a set of symmetry-adapted descriptors, and can therefore be applied to other lattice models 
whose order parameters are not known in advance. The main limitation lies in the descriptor set, which has to be chosen beforehand; replacing it by descriptors learned directly from the configurations, subject to the symmetry of the lattice, is a natural direction for further work.

Finally, a natural extension concerns the final step of the workflow, which at present still relies on human input: the analytical ansatz for each phase is inferred by inspection of the relaxed configurations. When the potential is known explicitly, this step could be automated by applying symbolic regression to the configurations belonging to each cluster, and the resulting candidate expressions could then be verified directly by substituting them into the potential and testing their stability under given conditions. The phase diagram would thus be constructed together with closed-form descriptions of the phases it contains.

\section*{Data and code availability}
The configurations obtained from simulated annealing,
together with the precomputed descriptors, are available at
Ref.~\cite{zenodo}. The implementations of the descriptor calculation and of the contrastive
and soft contrastive training pipelines are openly available at
Ref.~\cite{repo}. 

\section*{Acknowledgement}
\label{Sec:Acknowledgement}

We acknowledge the assistance provided by the  Operational Programme Johannes Amos Comenius of the Ministry of Education, Youth  and  Sport of the Czech Republic, within the frame of project Ferroic Multifunctionalities (FerrMion) [project No. CZ.02.01.01/00/22\_008/0004591], co-funded by the European Union. 

\appendix

\section{Encoder architecture}
\label{app:encoder}

The input configurations are sampled on a $8 \times 8$ grid with 4 channels corresponding to the 4 joint components of fields $\bp$ and $\bA$. Due to the image-like qualities of the input data, we naturally opted for a convolutional neural network as the encoder.

\begin{figure}[t]
\centering
\begin{tikzpicture}[
  font=\small,
  node distance=3.2mm,
  arrow/.style={-latex, thick, Gray!80!Black},
  layer/.style={draw=Gray!80!Black, thin, rounded corners, minimum height=6mm, minimum width=50mm, inner sep=5pt, align=center},
  input/.style={layer, draw=Goldenrod!80!Black, fill=Goldenrod!20},
  output/.style={layer, draw=Goldenrod!80!Black, fill=Goldenrod!20},
  dim/.style={text=Black},
  split/.style={layer, rectangle split, rectangle split part align={center,center}},
  conv/.style={split, draw=YellowGreen!80!Black, rectangle split parts=2, rectangle split part fill={YellowGreen!15, White}},
  fc/.style={split, draw=TealBlue!80!Black, rectangle split parts=2, rectangle split part fill={TealBlue!15, White}},
  pool/.style={layer, draw=Lavender!80!Black, fill=Lavender!15}
]

\node[input] (inp) {Input};
\node[dim, right=of inp] {$(8,8,4)$} ;

\node[conv, below=of inp] (conv1) {
    Conv $3\times3$  (circular pad)
    \nodepart{two}BN + SiLU
} ;
\node[dim, right=of conv1] {$(8,8,32)$} ;

\node[conv, below=of conv1] (conv2) {
    Conv $3\times3$ (zero pad)
    \nodepart{two}BN + SiLU
} ;
\node[dim, right=of conv2] {$(8,8,64)$} ;

\node[pool, below=of conv2] (avg) {
    Global Average Pooling
} ;
\node[dim, right=of avg] {$(64)$} ;

\node[fc, below=of avg] (fc1) {
    Fully connected 
    \nodepart{two}BN + SiLU + Dropout
} ;
\node[dim, right=of fc1] {$(32)$} ;

\node[fc, below=of fc1] (fc2) {
    Fully connected
    \nodepart{two}BN + SiLU + Dropout
} ;
\node[dim, right=of fc2] {$(16)$} ;

\node[fc, below=of fc2] (out) {
    Fully connected
    \nodepart{two} Normalization
} ;
\node[dim, right=of out] {$(2)$} ;

\node[output, below=of out] (emb) {Output};
\node[dim, right=of emb] {$(d)$} ;

\draw[arrow] (inp) -- (conv1);
\draw[arrow] (conv1) -- (conv2);
\draw[arrow] (conv2) -- (avg);
\draw[arrow] (avg) -- (fc1);
\draw[arrow] (fc1) -- (fc2);
\draw[arrow] (fc2) -- (out);
\draw[arrow] (out) -- (emb);

\end{tikzpicture}
\caption{Encoder architecture diagram. Each box specifies a layer of neurons including the signal processing.
The numbers in the brackets correspond to the output dimension of given layer.
Abbreviations are explained in the main text.}
\label{fig:encoder-arch}
\end{figure}
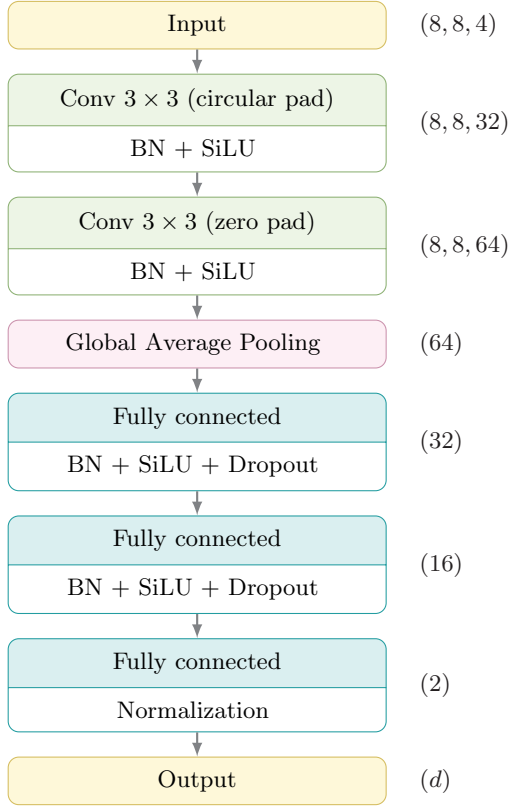

A diagram of the used architecture can be seen in Figure \ref{fig:encoder-arch}. We used two convolutional layers followed by averaged pooling and finally applying three fully connected layers. The convolutions were performed using $3\times3$ kernels with stride 1. The circular padding in the first layer respects the periodic boundary conditions. Each layer except the final one is followed by a batch normalization (BN) and SiLU activation. The fully connected layers are also regularized using dropouts. After the final layer, the embedding vectors are normalized. The dimensions of the layers are specified in the figure.

\section{Dimension of embeddings}\label{app:collapse}
We analyzed the dimensions of embeddings across both the classical and soft contrastive branches of  our workflow. The main motivation was an apparent dimensional collapse in the classical contrastive learning case for embedding dimension $D=3$, where the embeddings almost perfectly spanned a 2-dimensional hyperplane.

The main tool in this case was the principal component analysis (PCA), which performs a change of basis, so that the resulting axes align with the directions of greatest variance in the data. The transformed data is known as the principal components (PC). The variances associated with each PC are denoted $\lambda_i$, where $i \in \mathbb{N}$, so that the numbers are ordered in descending order, $\lambda_1 \geq \lambda_2 \geq \cdots \geq \lambda_N$.
We performed a parameter sweep by letting both encoders train with different initial conditions. We changed the target embedding dimension $D \in \{ 3, 4, 6, 8, 12\}$, while keeping other hyperparameters identical to the parameters of the runs used to obtain the results in Sec. \ref{Sec:Results}. For each of the dimensions, we repeated the training ten times.

\begin{figure}
    \centering
    \includegraphics[width=0.8\linewidth]{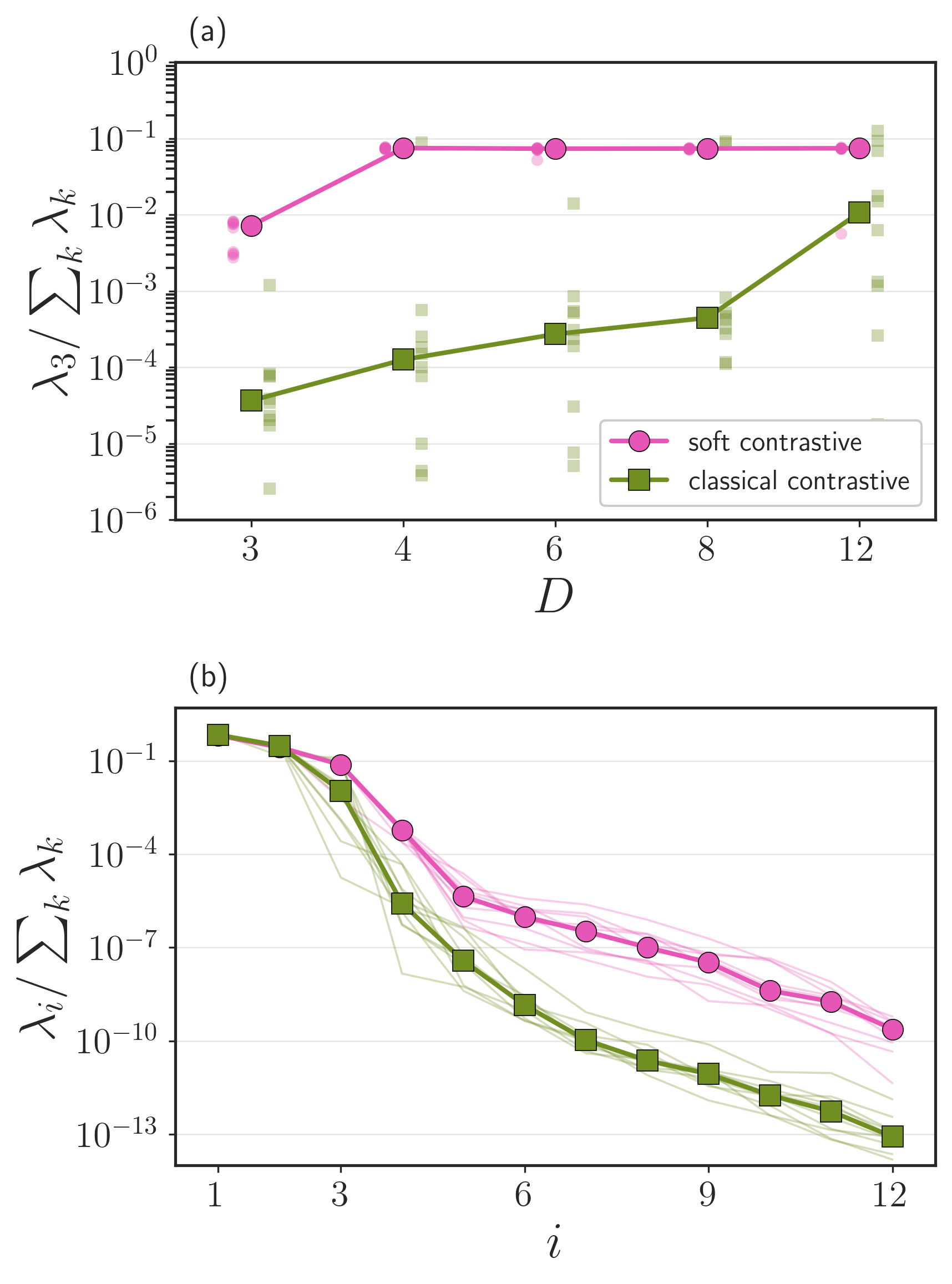}
    \caption{Results of PCA showing the explained variance ratios $\lambda_i/\sum_k\lambda_k$. Tests were performed across 10 independent training runs for both encoders. (a) Explained variance ratio of the third largest PC, $\lambda_3/\sum_k \lambda_k$, for varying embedding dimension $D$. (b) All explained variance ratios $\lambda_i/\sum_k \lambda_k$, where $i \in \{1, 2, \dots, 12\}$ for $D=12$. }
    \label{fig:third_direction}
\end{figure}

The results are given in Fig.~\ref{fig:third_direction}. Panel~(a) shows value of the explained variance ratio, $\lambda_3/\sum_k\lambda_k$, in logarithmic scale for each embedding dimension across both methods. The highlighted points are median values and transparent points display individual runs across different seeds.  We can observe a gap between the values of soft and classical contrastive methods, with the soft method having higher values of $\lambda_3/\sum_k\lambda_k$ by about two orders of magnitude for $D \leq 8$. For the soft contrastive branch, the values plateau above $D\geq 4$ at about $7.43\times10^{-2}$. Panel~(b) shows general $\lambda_i/\sum_k\lambda_k$, where $i \in \{1, 2, \dots, 12\}$, for $D=12$. For both methods, most of the variance is carried by the first two principal components PC1 and PC2, more precisely $99.996\%$ for classical contrastive and $99.278\%$ for soft contrastive. 
\vspace{2pt}

As for our choice of embedding dimensions, we opted for $D=3$ in the case of soft contrastive learning due to increased ability to show difference between the AFE A and AFE A' configurations. When we set the embedding dimension to $D=2$, the embeddings of the two AFE configurations overlapped, rendering the clustering algorithm unable to reliably distinguish them. On the other hand, the classical contrastive algorithm accurately found all the expected phases for $D=2$. For $D=3$, the classical embedding remained effectively two-dimensional, whereas the soft-contrastive one retained a small but consistently non-zero third direction, e.g. Fig.~\ref{fig:sft_ctr_result}. However, the PCA results show that at $D=3$ both sets of embeddings carry little variance in the third principal component, even in the seemingly non-degenerate result obtained in the soft-contrastive learning case.

\FloatBarrier
\end{document}